\documentstyle[12pt,psfig,draft,lscape,array,amssymb]{nature-ejb-new}

\def\simlt{\mathrel{\hbox{\rlap{\hbox{\lower4pt\hbox{$\sim$}}}\hbox{$<$}}}}
\def\simgt{\mathrel{\hbox{\rlap{\hbox{\lower4pt\hbox{$\sim$}}}\hbox{$>$}}}}

\def\ale{\mathrel{\hbox{\rlap{\hbox{\lower4pt\hbox{$\sim$}}}\hbox{$<$}}}}
\def\age{\mathrel{\hbox{\rlap{\hbox{\lower4pt\hbox{$\sim$}}}\hbox{$>$}}}}

\def\nodata{---}

\def\ra#1#2#3{#1$^{\rm h}$#2$^{\rm m}$#3$^{\rm s}$}
\def\dec#1#2#3{$#1^\circ#2'#3''$}

\def\spose#1{\hbox to 0pt{#1\hss}}
\newcommand\lsim{\mathrel{\spose{\lower 3pt\hbox{$\mathchar"218$}}
     \raise 2.0pt\hbox{$\mathchar"13C$}}}
\newcommand\gsim{\mathrel{\spose{\lower 3pt\hbox{$\mathchar"218$}}
     \raise 2.0pt\hbox{$\mathchar"13E$}}}

\begin{document} 

\title{\Large \bf A relativistic type Ibc supernova without a detected $\gamma-$ray burst}

\author{
A.~M.~Soderberg\affiliation[1]{Harvard-Smithsonian Center for Astrophysics, 60 Garden Street, MS-51, Cambridge, MA 02138, USA},
S.~Chakraborti\affiliation[2]{Tata Institute of Fundamental Research, Mumbai 400 005, India},
G.~Pignata\affiliation[3]{Departamento de Astronomi'a, Universidad de Chile, 
Casilla 36-D, Santiago, Chile},
R.~A.~Chevalier\affiliation[4]{University of Virginia, Department of Astronomy,
 PO Box 400325, Charlottesville, VA 22904, USA},
P.~Chandra\affiliation[5]{Royal Military College of Canada, Kingston, ON Canada},
A.~Ray\affiliationmark[2],
M.~H.~Wieringa\affiliation[6]{Australia Telescope National Facility, CSIRO, Epping 2121, Australia},
A.~Copete\affiliationmark[1],
V.~Chaplin\affiliationmark[7], 
V.~Connaughton\affiliation[7]{University of Alabama, Huntsville, AL, USA},
S.~D.~Barthelmy\affiliation[8]{NASA Goddard Space Flight Center, 
Greenbelt, MD 20771, USA},
M.~F.~Bietenholz\affiliation[9]{Department of Physics and Astronomy, York 
University, Toronto, ON M3J 1P3, Canada}$^,$\affiliation[10]{Hartebeestehoek 
Radio Observatory, PO Box 443, Krugersdorp, 1740, South Africa},
N.~Chugai\affiliation[11]{Institute of Astronomy, RAS, Pyatnitskaya 48,
Moscow 119017, Russia},
M.~D.~Stritzinger\affiliation[12]{Las Campanas Observatory, Carnegie Observatories, Casilla 601, La Serena, Chile}$^,$\affiliation[13]{Dark Cosmology Centre, Niels Bohr Institute, University of Copenhagen, Copenhagen, Denmark},
M.~Hamuy\affiliationmark[3],
C.~Fransson\affiliation[14]{Department of Astronomy, Stockholm University, AlbaNova, SE-106 91 Stockholm, Sweden},
O.~Fox\affiliationmark[4],
E.~M.~Levesque\affiliationmark[1]$^,$\affiliation[15]{Institute for
Astronomy, University of Hawaii, 2680 Woodlawn Dr., Honolulu, HI
96822, USA},
J.~E.~Grindlay\affiliationmark[1],
P. Challis\affiliationmark[1],
R.~J.~Foley\affiliationmark[1],
R.~P.~Kirshner\affiliationmark[1],
P.~A.~Milne\affiliation[16]{Steward Observatory, University of Arizona, 933 
North Cherry Avenue, Tucson, AZ 85721, USA},
M.~A.~P.~Torres\affiliationmark[1]
}
\date{\today}{}
\headertitle{A Relativistic SN}
\mainauthor{Soderberg et al.}

\summary{Long duration gamma-ray bursts (GRBs) mark\cite{wb06} the
explosive death of some massive stars and are a rare sub-class of Type
Ibc supernovae (SNe Ibc).  They are distinguished by the production of
an energetic and collimated relativistic outflow powered\cite{mwh01}
by a central engine (an accreting black hole or neutron star).
Observationally, this outflow is manifested\cite{pir99} in the pulse
of gamma-rays and a long-lived radio afterglow.  To date, central
engine-driven SNe have been discovered exclusively through their
gamma-ray emission, yet it is expected\cite{pl98} that a larger
population goes undetected due to limited satellite sensitivity or
beaming of the collimated emission away from our line-of-sight. In
this framework, the recovery of undetected GRBs may be possible
through radio searches\cite{bkf+03,snb+06} for SNe Ibc with
relativistic outflows.  Here we report the discovery of luminous radio
emission from the seemingly ordinary Type Ibc SN\,2009bb, which
requires a substantial relativistic outflow powered by a central
engine.  The lack of a coincident GRB makes SN\,2009bb the first
engine-driven SN discovered without a detected gamma-ray signal.  A
comparison with our extensive radio survey of SNe Ibc reveals that the
fraction harboring central engines is low, $\sim 1\%$, measured
independently from, but consistent with, the inferred\cite{skn+06}
rate of nearby GRBs.  Our study demonstrates that upcoming optical and
radio surveys will soon rival gamma-ray satellites in pinpointing the
nearest engine-driven SNe.  A similar result for a different supernova
is reported\cite{par09} independently.}

\maketitle


On 2009 Mar 21.1 UT, the Chilean Automatic Supernova Search Program
(CHASE; Ref.~\pcite{pma+09}) discovered\cite{pmh+09} a bright optical
transient through repeated imaging of the nearby spiral galaxy
NGC\,3278 at a distance, $d\approx 40$ Mpc. The new object was offset
22 arcsec (4.2 kpc) from the center of the galaxy and located within
its star-forming disk.  Optical spectroscopy obtained on Mar 28.1 UT
revealed\cite{spm+09} that the transient was a young Type Ibc SN
(hereafter SN\,2009bb) lacking evidence for Hydrogen in the explosion
debris.  Based on the previous non-detection of SN\,2009bb on Mar 19.2
UT, we tightly constrain the SN explosion date to be Mar $19\pm 1$ UT
(see Suppl.~Info.).

Using the Very Large Array (VLA) on Apr 5.2 UT, we discovered a
coincident radio counterpart at $\alpha$(J2000)=\ra{10}{31}{33.87} and
$\delta$(J2000)=\dec{-39}{57}{30.1} ($\pm 0.7$ arcsec in each
coordinate) and with a flux density, $F_{\nu}=24.53\pm 0.06$ mJy, at
frequency, $\nu=8.46$ GHz.  This corresponds to a spectral radio
luminosity of $L_{\nu}\approx 5\times 10^{28}~\rm erg~s^{-1}~Hz^{-1}$
at $\Delta t\approx 17$ days after explosion, more luminous than any
other SN Ibc observed\cite{bkf+03,snb+06,ams07} on a comparable
timescale. Instead, the radio properties of SN\,2009bb are consistent
with the sample of nearby ($z\lesssim 0.1$) GRBs, observed to
consistently yield\cite{skn+06} lower relativistic energies than
``classic" GRBs preferentially discovered at larger distances.
Further VLA observations of SN\,2009bb revealed a power-law flux
decay, $F_{\nu,8.46~{\rm GHz}}\approx
 t^{-1.4}$, in line with the radio afterglow evolution
 seen\cite{kfw+98} for the nearest
 gamma-ray burst, GRB\,980425 at a similar distance of $d\approx 38$
 Mpc (Figure~1).

Unlike the optical emission from SNe which traces only the slowest
explosion debris, radio observations uniquely probe\cite{c98} the {\it
fastest} ejecta as the expanding blastwave (velocity, $v$) shocks and
accelerates electrons in amplified magnetic fields.  The resulting
synchrotron emission is suppressed by self-absorption (SSA) producing
a low frequency radio turn-over that defines the spectral peak
frequency, $\nu_p$.  Combining our observations from the VLA and the
Giant Meterwave Radio Telescope (GMRT), the radio spectra of
SN\,2009bb are well described by an SSA model across multiple epochs
(Figure~2). From our earliest spectrum on Apr 8 UT
($\Delta t\approx 20$ days), we infer $\nu_p\approx 6$ GHz and a
spectral peak luminosity, $L_{\nu,p}\approx 3.6\times 10^{28}~\rm
erg~s^{-1}~Hz^{-1}$.

Making the conservative assumption that the energy of the radio
emitting material is partitioned equally into accelerating electrons
and amplifying magnetic fields (equipartition), the properties of the
SSA radio spectrum enable\cite{kfw+98,c98} a robust estimate of the
blastwave radius, $R\approx 2.9\times 10^{16} (L_{\nu,p}/10^{28}~{\rm
erg~s^{-1}~Hz^{-1}})^{9/19} (\nu_p/5~{\rm GHz})^{-1}$ cm.  Luminous
synchrotron sources with a low spectral peak frequency thus require
larger sizes (Figure~3).  For SN\,2009bb, we infer
$R\approx 4.4\times 10^{16}$ cm at $\Delta t\approx 20$ days and thus
the mean expansion velocity is $R/\Delta t=0.85\pm 0.02c$, where $c$
is the speed of light.  The transverse expansion speed, $\Gamma\beta c
= R/\Delta t$ indicates that the blastwave is relativistic, $\Gamma
\gtrsim 1.3$, at this time [bulk Lorentz factor
$\Gamma=(1-\beta^2)^{-1/2}$ with $\beta=v/c$].  This is a lower limit
on the initial velocity since the radio evolution indicates that the
blastwave decelerated early on.  We further find that the radio
emission requires a minimum energy, $E=(1.3\pm 0.1)\times 10^{49}$
erg, coupled to the relativistic outflow and comparable to the values
inferred\cite{skn+06,kfw+98,lc99,skb+04b} from the radio afterglows of
nearby GRBs (see Figure~4).

These conclusions are robust; the blastwave velocity is
insensitive\cite{c98,cf06} to deviations from equipartition while the
relativistic energy can only be higher\cite{r94}. In view of these
constraints, we note that shock-acceleration in some SNe Ibc
may\cite{tmm+01} couple a minute fraction ($\lesssim 0.01\%$) of the
total energy, $E_{\rm tot}$, to material with a trans-relativistic
velocity.  However, this scenario would require an exceedingly high
total energy for SN\,2009bb, $E_{\rm tot}\gtrsim 10^{53}$ erg, a
factor of $10^2$ higher than the total explosion energies inferred for
SNe Ibc.  We conclude that the energetic and relativistic outflow from
SN\,2009bb was powered by another energy reservoir, a central engine.
To date, engine-driven SNe have been discovered exclusively through
their gamma-ray emission, making SN\,2009bb the first to be identified
by its long-wavelength signal.

Motivated by our discovery of an engine-driven relativistic outflow,
we searched for a gamma-ray counterpart in temporal and spatial
coincidence with SN\,2009bb.  During our bracketed explosion date
estimate, the all-sky Interplanetary Network (IPN;
Ref.~\pcite{hcm+09}) of high energy satellites did not detect a
coincident GRB (see Suppl.~Info.).  Based on the IPN sensitivity and
detection efficiency, we place an upper limit on the gamma-ray fluence
of $F_{\gamma} < 5\times 10^{-6}~\rm erg~cm^{-2}$, corresponding to an
energy of $E_{\gamma} < 10^{48}$ erg (band, 25-150 keV) if the
relativistic outflow was isotropic.  This limit is a factor of two
higher than the isotropic-equivalent $E_{\gamma}$
observed\cite{paa+00} from GRB\,980425, and thus it is possible that
SN\,2009bb gave rise to a similar (albeit undetected) signal. At the
same time, these limits cannot exclude scenarios in which the SN (i)
powered a GRB directed away from our line-of-sight, or (ii) did not
produce any gamma-rays.  SN\,2009bb observationally demonstrates the
limitation of using gamma-ray satellites as a primary tool to identify
nearby engine-driven explosions.

In this context, we note that our VLA observations of SN\,2009bb were
obtained as part of an extensive radio survey of 143
optically-discovered local SNe Ibc designed to recover relativistic
SNe without detected gamma-ray counterparts.  This systematic radio
study revealed\cite{skn+06,bkf+03,snb+06,ams07} no additional
relativistic SNe, instead indicating typical maximum blastwave
velocities for SNe Ibc of $\beta\approx 0.1$
(Figures~3 and 4).  From this
sample, we therefore estimate the fraction of engine-driven SNe to be
just $0.7^{+1.6}_{-0.6}$\% ($1\sigma$).  This is measured
independently from, and yet consistent with, the fraction inferred
from the relative rates\cite{skn+06,cet99,dsr+04} of nearby GRBs and
all SNe Ibc, $1.4^{+5.0}_{-0.3}$\% (1$\sigma$).
Our long-term study thus confirms that engine-driven SNe are uncommon.

The infrequency of relativistic outflows among massive star explosions
implies that their progenitor stars share an essential and rare
physical property. Observations of the explosion environment may offer
unique clues.  On a local scale ($\lesssim 1$ pc), where the
environment was shaped directly by the evolution of SN\,2009bb's
progenitor star, we find evidence for a pre-explosion mass loss rate
of $\dot{M}=(2.0\pm 0.2)\times 10^{-6}~\rm M_{\odot}~yr^{-1}$,
consistent with the wide distribution of values
inferred\cite{lc99,skb+04b,skn+06} for nearby GRBs.  However, the
large-scale ($\gtrsim 1$ kpc) environmental properties
differ\cite{lsf+09} from those of nearby GRB host galaxies, showing
evidence for a super-solar metallicity that exceeds the
proposed\cite{wh06,mkk+08} cut-off for relativistic explosions.  We
conclude that there is a broad diversity in the environments of
engine-driven explosions, and therefore host galaxy properties alone
cannot be used to discriminate between ordinary and engine-driven SNe.

With the advent\cite{lkd+09,kab+02} of wide-field optical surveys
(e.g., Palomar Transient Factory, Pan-STARRS) the discovery rate of
young, local SNe Ibc will effectively quadruple over the next 3-5
years (see Suppl. Info). Coupled with the ten-fold increase in the
sensitivity of the Expanded Very Large Array (expected 2010;
Ref.~\pcite{pnb04}), relativistic SNe will be uncovered at an
increased rate of $\sim 1$ per year within $d\lesssim 200$ Mpc.  This
is $\sim 3$ times higher than the rate at which nearby GRBs are
discovered with current gamma-ray satellites. Thus, while such
explosions have historically been found through their gamma-ray
emission, long-wavelength surveys will soon provide a more powerful
tool to pinpoint the nearest engine-driven supernovae.

\bibliographystyle{nature-pap}

\noindent 
{\bf \large Supplementary Information} is linked to the online version of the
paper at www.nature.com/nature.

\begin{acknowledge} The VLA is operated by the National Radio
Astronomy Observatory, a facility of the National Science Foundation
operated under cooperative agreement by Associated Universities, Inc.
GMRT is run by the National Centre for Radio Astrophysics of the Tata
Institute of Fundamental Research.  This research has made use of the
NASA/IPAC Extragalactic Database (NED) which is operated by the Jet
Propulsion Laboratory, California Institute of Technology, under
contract with the National Aeronautics and Space Administration.
A.M.S.  and O.F. acknowledge support by NASA through Hubble and GSRP
grants, respectively.  E.M.L.  is funded through a Ford
Foundation Predoctoral Fellowship. R.J.F. is a Clay fellow.
R.A.C. and R.P.K. acknowledge support through NASA and NSF grants.  G.P. and
G.P. and M.H. acknowledge support from FONDECYT, Iniciativa Cientifica
Milenio, FONDAP, and CONICYT. A.R. and S.C. are funded by an 11th Five Year Plan Project.
\end{acknowledge}

\bigskip
\noindent
All authors contributed extensively to the work presented in this
paper.  Reprints and permissions information is available at
npg.nature.com/reprintsandpermissions.  Correspondence should be
addressed to A.~M.~Soderberg (e-mail: asoderberg@cfa.harvard.edu).

\clearpage

\bigskip \noindent {\bf Figure~1:~ Radio observations of the nearest
massive star explosions.}  The 8.46 GHz radio emission from SN\,2009bb
(red) is more luminous than any of the other 142 local ($d\lesssim
200$ Mpc) SNe Ibc observed (Ref.\pcite{ams07} and references within)
to date on a comparable timescale ($\Delta t\lesssim 100$ days), and
is consistent\cite{skn+06,kfw+98,skb+04b} with the radio afterglow
luminosities of the nearest GRBs discovered through their gamma-ray
signal within a similar volume (black).  Local SNe Ibc with
well-studied radio emission (grey) exhibit lower luminosities and peak
at later times, indicating smaller sizes and lower mean expansion
velocities.  The radio emission from most local SNe Ibc is below our
current detection threshold; we include them here as upper limits
($3\sigma$; grey triangles).

\bigskip\noindent {\bf Figure~2:~ Synchrotron self-absorption model
fits to the SN\,2009bb radio spectra.}  The radio emission from GRBs
and SNe Ibc is suppressed\cite{c98,cf06} at low frequencies by SSA,
defining the spectral peak frequency.  The spectral shape below and
above the peak is characterized as $F_{\nu}\propto \nu^{5/2}$ and
$F_{\nu}\propto \nu^{-(p-1)/2}$, respectively, where $p$ is the
power-law index of the relativistic electron energy distribution above
a minimum Lorentz factor, $\gamma_m$.  Our multi-frequency radio
observations of SN\,2009bb taken with the VLA and GMRT (see
Suppl.~Info.) on Apr 8, May 10, Jun 6-10, and Aug 8-11 UT ($\Delta
t\approx 20,~52,~81,$ and 145 days) are well described by a standard
SSA spectrum, with $\nu_p\approx 6,~3,~1,$ and 0.8 GHz, and peak flux
densities of $F_{\nu,p}\approx 19,~15,~13,$ and 11 mJy,
respectively. As the blastwave expands the shocked material becomes
optically thin, causing $\nu_p$ to cascade\cite{cf06} to lower
frequencies with time.  The optically thin spectral index is
constrained to roughly $F_{\nu}\approx \nu^{-1}$ which implies
$p\approx 3$, in line\cite{cf06} with other radio SNe Ibc.  Error bars
are $1\sigma$.

\bigskip \noindent {\bf Figure~3: ~ Radio properties of the nearest
massive star explosions directly reveal the blastwave velocities.}  We
compare the peak radio luminosities for SNe Ibc (red) and nearby GRBs
($z\lesssim 0.1$; blue squares) as observed at the spectral peak
frequency, $\nu_p$, and at time $\Delta t$.  These observed properties
are tightly related\cite{c98} to the blastwave radius. The average
velocities are reasonably estimated as $R/{\Delta t}$ (dashed grey
lines).  For SNe Ibc we infer typical velocities of $R/\Delta t\approx
0.1$c, while SN\,2009bb (star) and the nearest GRBs show $R/\Delta
t\approx c$.   Error bars are $1\sigma$.

\clearpage

\bigskip \noindent {\bf Figure~4: ~ Blastwave velocity and energy for
massive star explosions.}  We compare the blastwave velocities and
inferred energies for the well-studied radio SNe Ibc and nearby GRBs
included in Figure~3 with classic GRBs.  Assuming
equipartition of energy between electrons ($\epsilon_e$) and magnetic
fields ($\epsilon_B$) as $\epsilon_e=\epsilon_B=0.33$ which accounts for
equal energy in shocked protons, the total
internal energy of the radio emitting source is $E\equiv E_{\rm
min}/{\epsilon_B}$ where the minimum energy, $E_{\rm min}$, is
derived\cite{cf06} from the energy density in magnetic fields, $E_{\rm
min}\approx 8.3\times 10^{46} (f/0.5) (B/{\rm G})^2 (R/10^{16}~{\rm
cm})^3~\rm erg$, here $B\approx 0.43 (\epsilon_e/\epsilon_B)^{-4/19}
(f/0.5)^{-4/19} (L_{\nu,p}/10^{28}~{\rm erg~s^{-1}~Hz^{-1}})^{-2/19}
(\nu_p/5~{\rm GHz})$ G and $f=0.5$ is the fraction of the spherical
volume occupied by the radio-emitting region.  For radio SNe Ibc (red;
Ref.~\pcite{ams07} and references therein), the average velocities are
reasonably estimated as $\beta=R/{c\Delta t}$ since the bulk flow
is\cite{cf06} expanding freely. The radio properties of these objects
imply typical values of $\beta\approx 0.1$ and $E\approx 10^{47}$ erg.
Adopting the same framework for SN\,2009bb (yellow star) we find
$R/\Delta t\approx 0.9c$ and $E\approx 10^{49}$ erg and note that the
inclusion of relativistic effects (e.g., photon arrival time) would
not change these results significantly as shown by Ref.~\pcite{kfw+98}
and \pcite{lc99} for GRB\,980425/SN\,1998bw.  The nearest GRBs
(blue squares) tend to be trans-relativistic and some decelerate rapidly,
so we adopt the inferred\cite{skn+06,kfw+98,lc99,skb+04b} blastwave
velocities and energies from the earliest available radio data, at
$\Delta t\lesssim 5$ days.  Likewise, the inferred velocity of
SN\,2009bb at $\Delta t\approx 20$ days, is a strict lower limit since
there is evidence that the blastwave decelerated early on (see
Suppl.~Info.).  Finally, classic GRBs (blue circles) in the
decelerating Blandford-McKee phase\cite{bm76} have $\Gamma\propto
t^{-3/8}$ so we conservatively estimate their blastwave velocity at
$\Delta t=1$ day according to $\Gamma\propto t^{-3/8}$ and adopt the
beaming corrected blastwave energies from afterglow modeling
(Ref.~\pcite{snb+06} and references therein).   Error bars
are $1\sigma$.

\clearpage

\begin{figure}
\centerline{\psfig{file=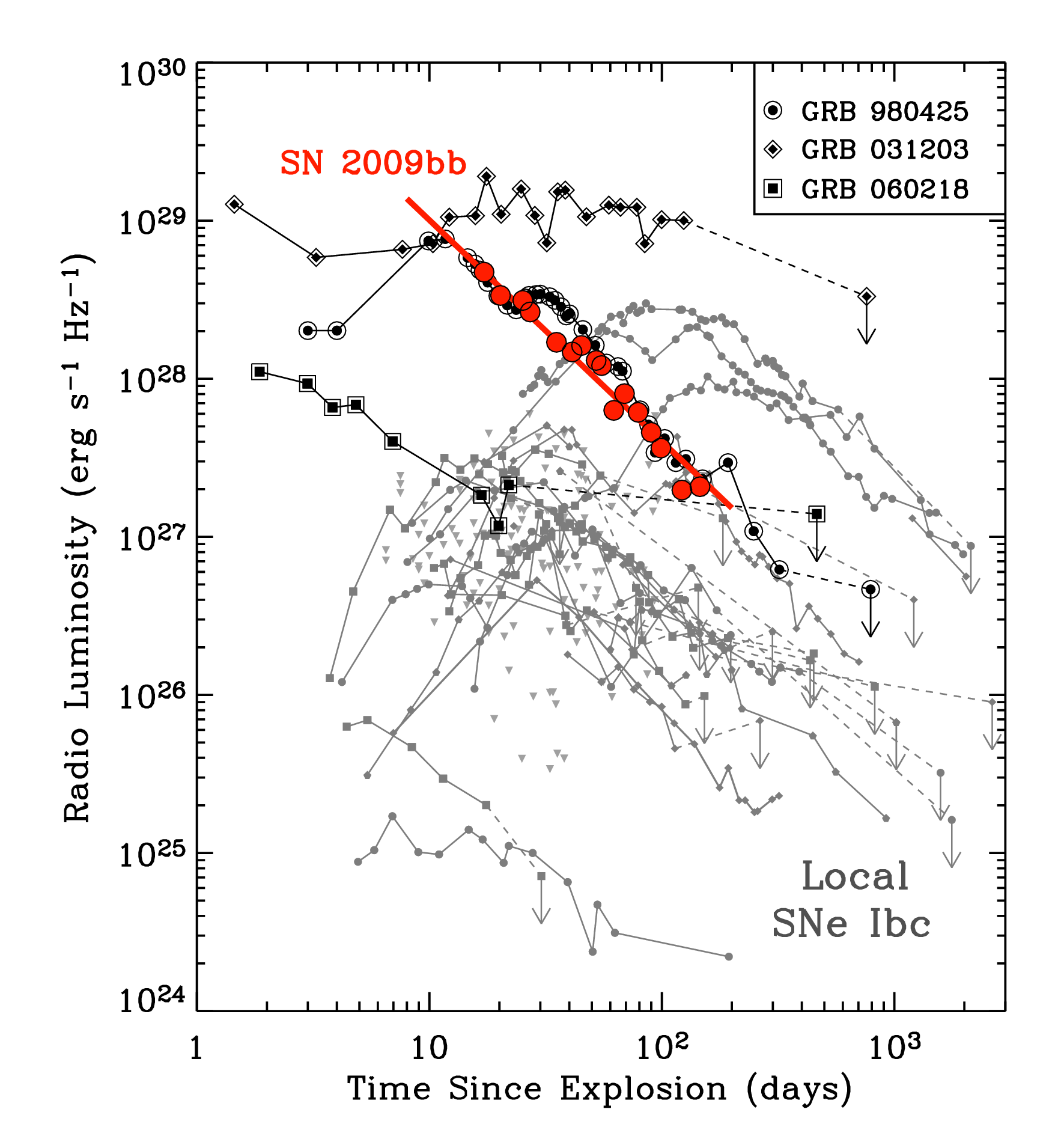,width=8in,angle=0}}
\caption[]{}
\label{fig:lum_limits_09bb}
\end{figure}

\clearpage

\begin{figure}
\centerline{\psfig{file=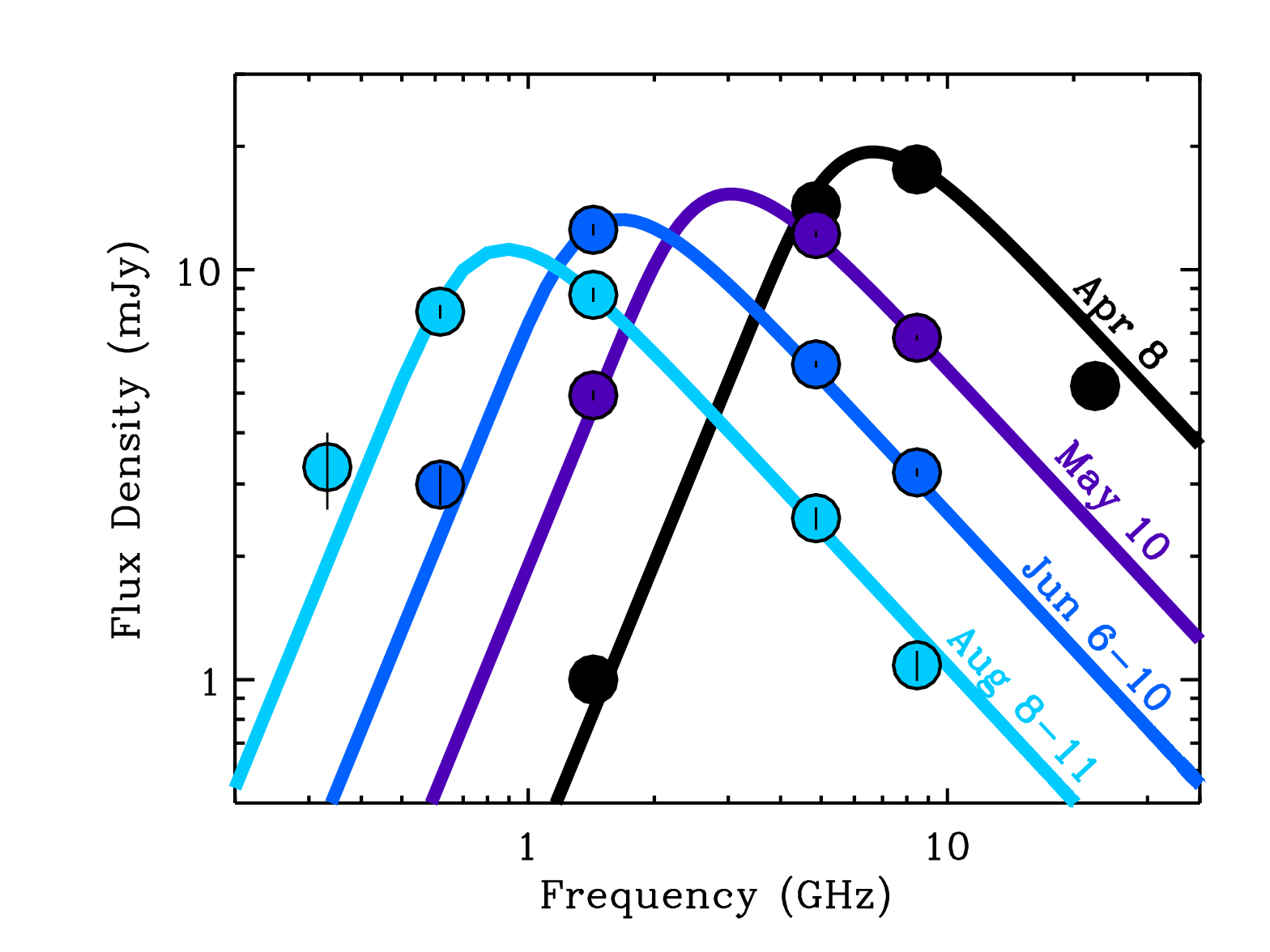,width=8in,angle=0}}
\caption[]{}
\label{fig:spectrum}
\end{figure}

\clearpage

\begin{figure}
\centerline{\psfig{file=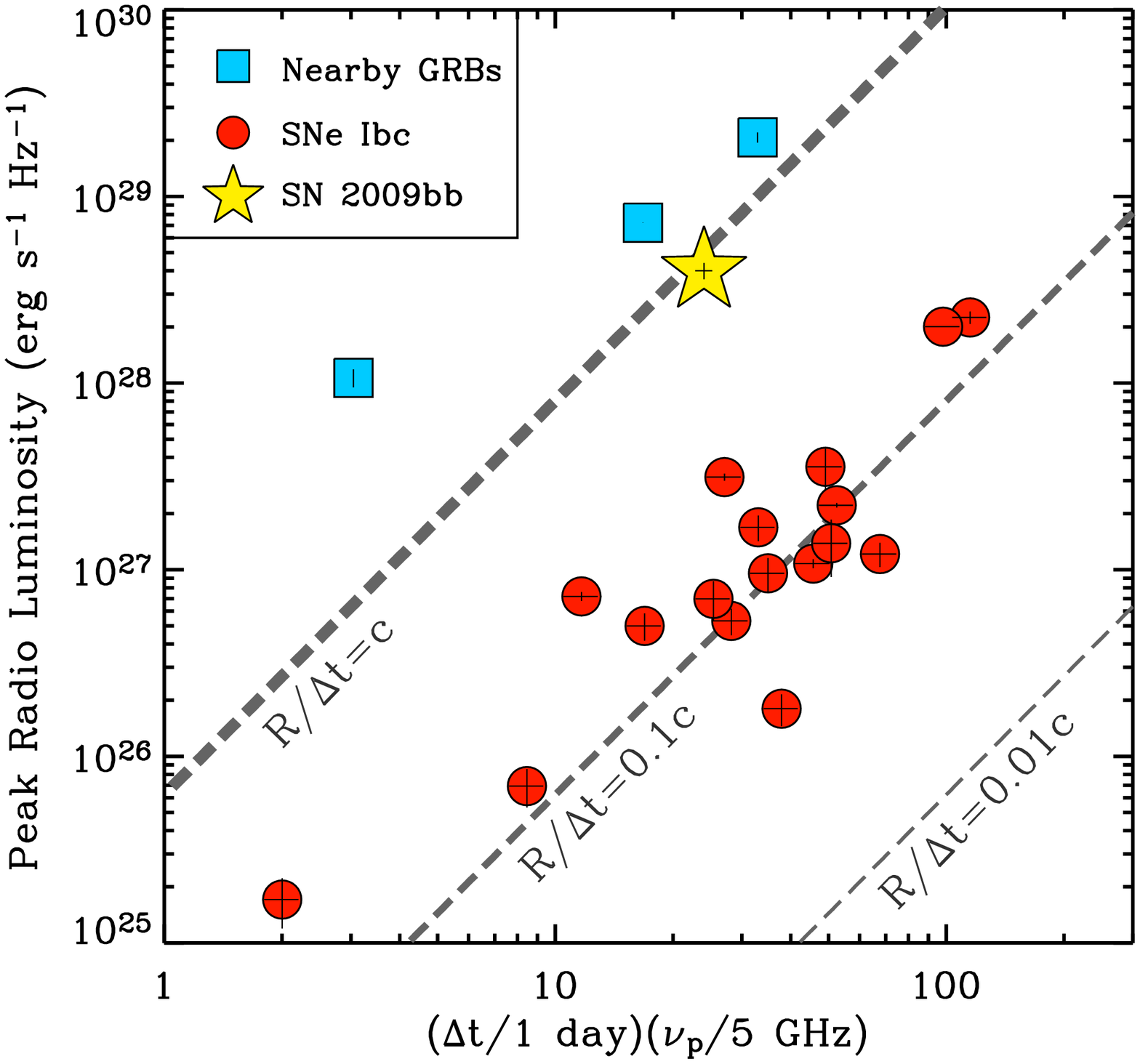,width=8in,angle=0}}
\caption[]{}
\label{fig:L_t_09bb}
\end{figure}

\clearpage

\begin{figure}
\centerline{\psfig{file=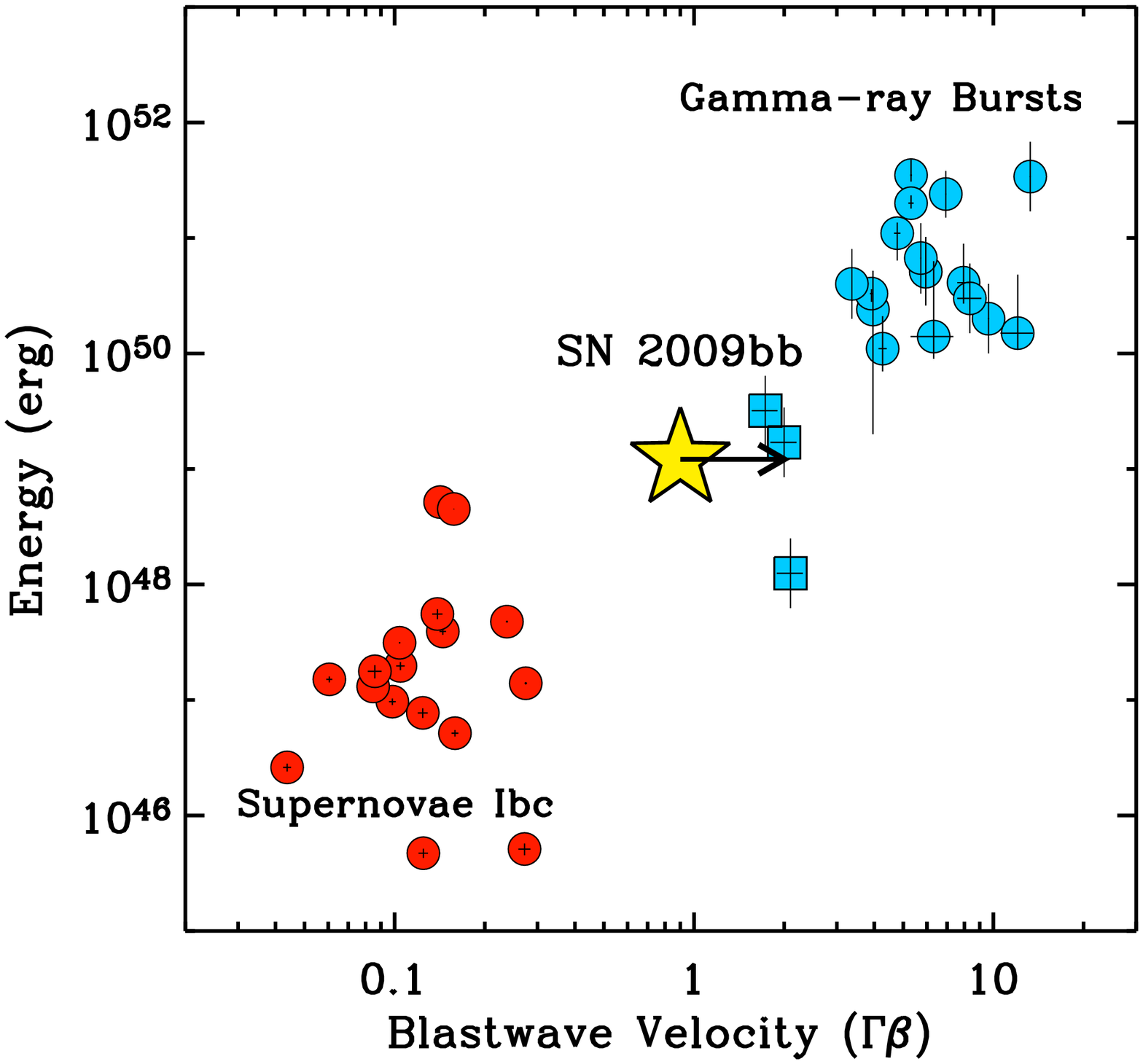,width=8in,angle=0}}
\caption[]{}
\label{fig:ev_sn2009bb}
\end{figure}

\clearpage

\begin{center}
{\Large SUPPLEMENTARY INFORMATION} \\
{\large \bf A Relativistic Supernova Powered by a Central Engine but Lacking a Detected $\gamma-$ray Burst}
\end{center}

\headertitle{Suppl. Info.}
\mainauthor{Soderberg et al.}

\section{Explosion Date}
\label{sec:t0}

The Chilean Automatic Supernova Search Program (CHASE;
Ref.~\pcite{pma+09}) and the Carnegie Supernova Project (CSP;
Ref.~\pcite{hfm+06}) initiated a dedicated optical and near-IR
follow-up campaign for SN\,2009bb beginning immediately after
discovery.  Using a combination of pre- and post-explosion data
collected by CHASE and CSP a well-sampled optical light-curve has been
constructed (G.~P., M.~D.~S. \& M.~A.~H. {\it in prep}).  Fitting a
expanding fireball model, $L_{\rm opt}\propto t^2$, to the rising
component of the optical light-curve implies an explosion date of
$t_0\approx$ Mar 18.4 UT $\pm$ 0.6 days.  While this fit enables a
first estimate of the explosion date, it assumes a constant velocity.
Realistically some deceleration is expected, especially for fast
expanding SNe like SN\,2009bb and SN\,1998bw so it is reasonable that the
explosion time should be somewhat later that the fireball fit.  Along this
line, we match the early evolution to that observed\cite{gvv+98} for
SN\,1998bw and find a slightly later explosion date, $t_0\approx$ Mar
20 UT.  We adopt an intermediate value of $t_0=$ Mar 19$\pm 1$ UT
throughout the paper.  Finally, we note that the inferred early
discovery is consistent with the observed epoch of maximum light, Apr
1 UT, occurring two weeks after our estimated explosion date and in
line with the observed rise times for other SNe Ibc (G.~P.,
M.~D.~S. \& M.~A.~H. {\it in prep}).

\section{Radio Observations}
\label{sec:radio}

\subsection{Very Large Array observations}

\begin{figure}
\centerline{\psfig{file=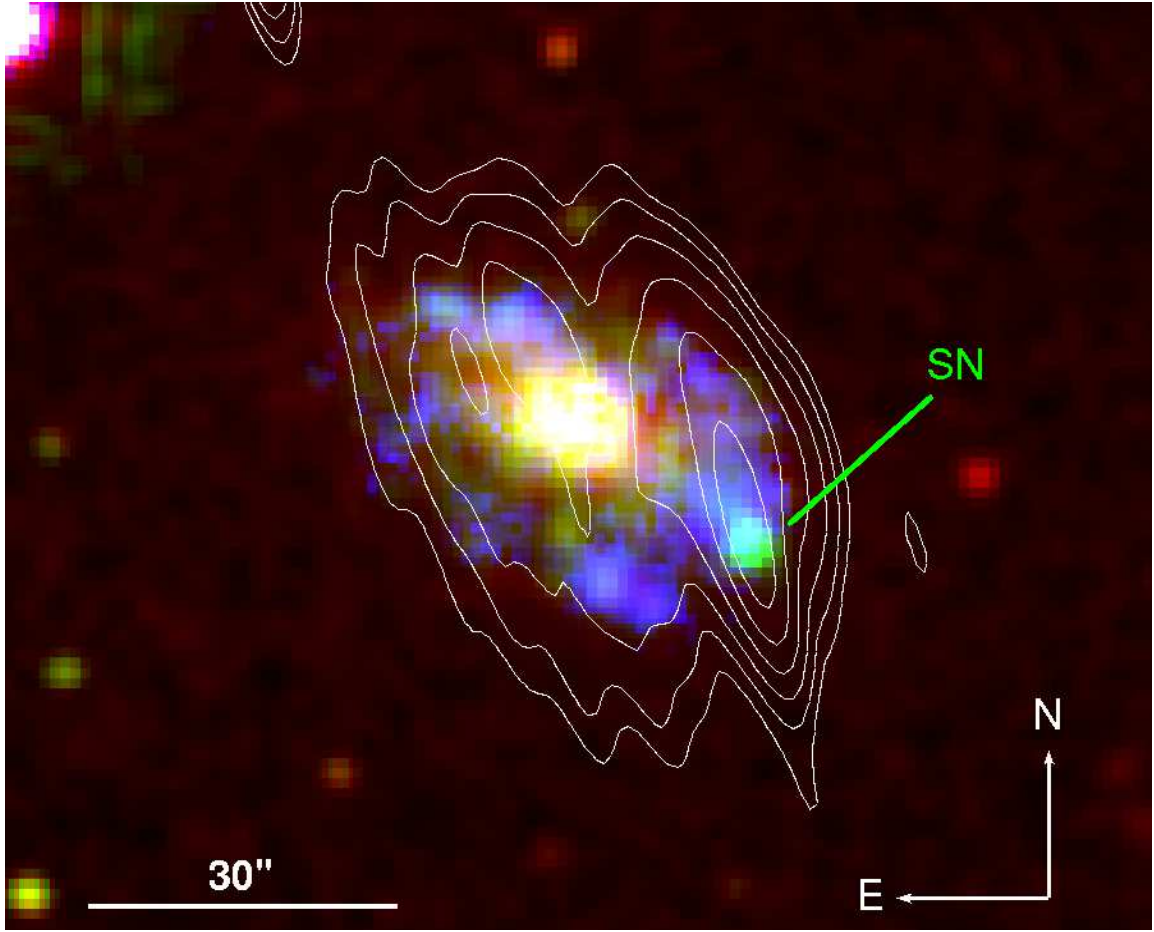,width=6in,angle=0}}
\caption[]{\small 
A composite image of the
host galaxy of SN\,2009bb, NGC\,3278, constructed from 2MASS
near-IR (red), {\it Swift}/UVOT optical (V-band, green) and UV
(UVW2-filter, blue) images.  Contours defining the structure of the
diffuse radio (1.43 GHz) emission as observed with the VLA in
B-configuration are over-plotted and trace the regions of
star-formation in the galaxy.  The SN is clearly detected in the radio
and optical images (labeled) and lies within a strongly star-forming
region of the galaxy.}
\label{fig:ds9_crop} 
\end{figure}

We observed SN\,2009bb with the Very Large Array (VLA) on many epochs
spanning Apr 5.2 to Jun 6.0 UT.  Data were collected at four
frequencies, $\nu_{\rm obs}=1.43, 4.86, 8.46$ and 22.5 GHz.  All VLA
observations were obtained in the standard continuum mode with
$2\times 50$ MHz bands.  At 22.5 GHz we included referenced pointing
scans to correct for the systematic 10-20 arcsec pointing errors of
the VLA antennas.  We used 3C286 (J1331+305) for flux calibration,
while phase referencing was performed against calibrator J1036-3744.
Data were reduced using standard packages within the Astronomical
Image Processing System (AIPS).  Flux density measurements were
obtained by fitting a Gaussian model to the radio emission.  For
observations at 1.43 GHz we restricted the UV-range
to reduce contamination from diffuse host galaxy emission from the
strongly star-formation region in which the SN resides
(Figure~\ref{fig:ds9_crop}).  The resulting flux density measurements
are listed in Table~\ref{tab:vla} and shown in
Figure~\ref{fig:lightcurves}.

\subsection{Giant Meterwave Radio Telescope Observations}

We observed the SN 2009bb with the Giant Meterwave Radio Telescope
(GMRT) on several epochs spanning May 31 and Aug 10 UT at effective
frequencies of 1287, 617, and 332 MHz.  We used a $2\times 16$ MHz
bandwidth each divided into 128 frequency channels.  The duration of
the observations was two hours each, except for the 332 MHz observation
which was 4 hrs in duration.  We used 3C147 and 3C286 for flux and
bandpass calibration and J1057-245, J1107-448, and J1018-317 to
monitor the phase at 1287, 617, and 332 MHz, respectively.

The GMRT data were reduced and analyzed with AIPS.  To remove
contamination from diffuse host galaxy emission, we removed the short
baselines 0--5 k$\lambda$ and 0--3 k$\lambda$ at 1287 and 617 MHz,
respectively.  At 332 MHz the host galaxy emission dominates that of
the SN at the SN position.  To estimate the flux density of the
diffuse emission in the beam of the SN, we first construct a host
galaxy map at 617 MHz and measure an integrated flux density of
$F_{617~\rm MHz}=113.5\pm 5.7$ mJy after removal of the SN emission.
 We convolve this map down to the beam size of our 332 MHz map, using
 the AIPS task CONVL. The extended emission at the SN site is then
 scaled to 332 MHz using the spectral
 index ($F_{\nu}\propto \nu^{-0.87}$) of the host galaxy
 emission as measured between 1287 and 617 MHz.
 The 332 MHz map is then 
 re-gridded to match the convolved 617 MHz map using the AIPS
 task HGEOM. The SN flux at the 332 MHz is then determined as the
 excess observed flux in the synthesized beam over and above the
 estimated extended emission at the SN position.  Our resulting flux
 densities for all GMRT epochs are listed in Table~\ref{tab:gmrt} and
 we note that the error in the SN flux density at 332 MHz is dominated
 by the uncertainty in determining the extended
 emission at 617 MHz.

\section{Radio Temporal Evolution}

The temporal and spectral evolution of the synchrotron emission is
determined by the dynamics, the density profile of the ejecta, and the
properties of the circumstellar medium.  Young, non-relativistic SNe
expand freely since their ejecta are largely undecelerated while GRBs
expand\cite{bm76} according to the Blandford-McKee solution for
relativistic, decelerated blastwaves.  Both SNe and GRBs ultimately
enter a phase of non-relativistic adiabatic expansion, the Sedov
phase, when the ejecta have fully decelerated and the transverse
velocity falls well below $\Gamma\beta \lesssim 1$. Trans-relativistic
blastwaves, for which $\Gamma\beta \sim$ a few, e.g.,
SN\,1998bw/GRB\,980425, bridge these three hydrodynamic regimes and
the flux density evolution depends on the structure of the ejecta.

\begin{figure}
\centerline{\psfig{file=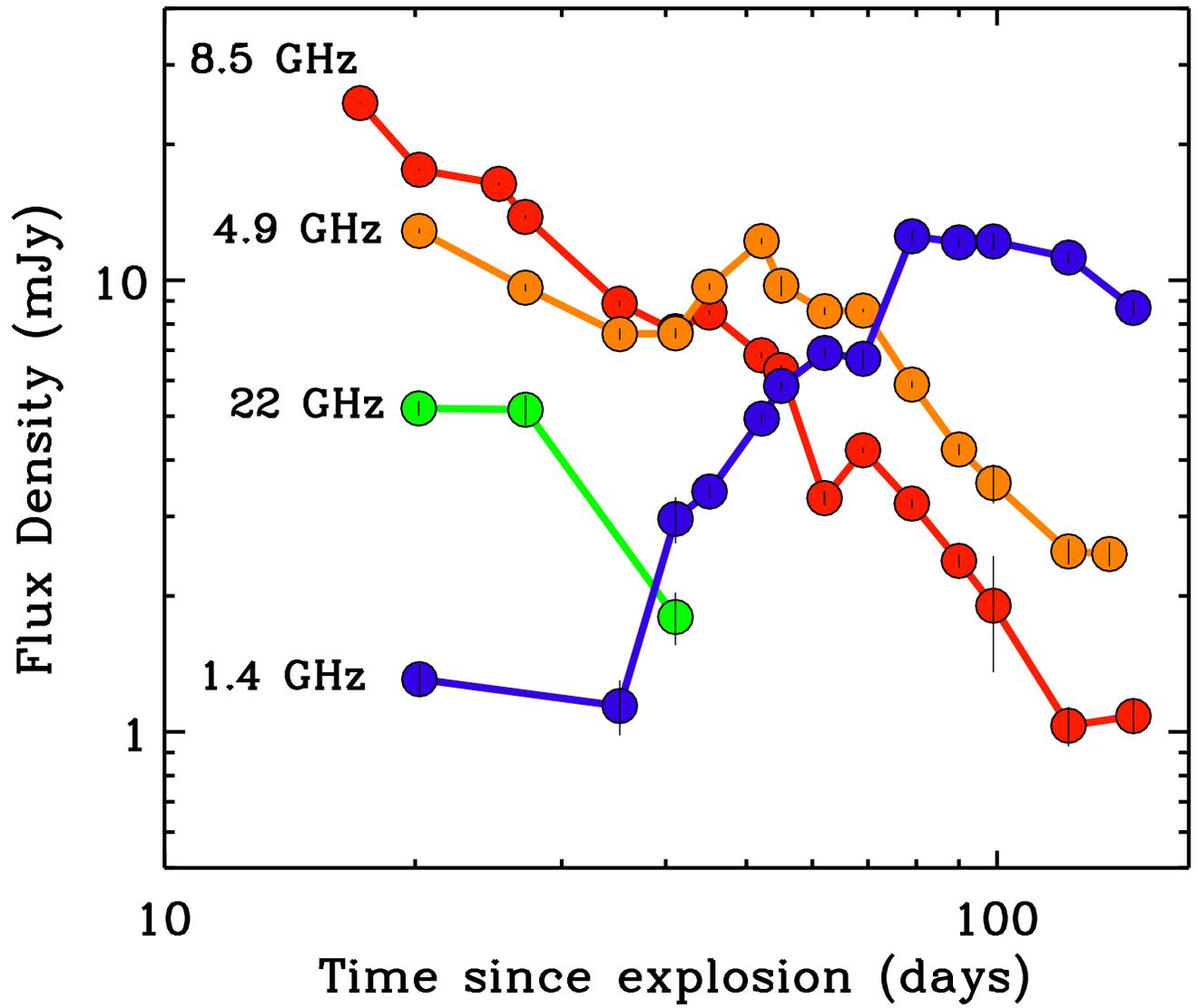,width=8in,angle=0}}
\caption[]{\small Radio light-curves for SN\,2009bb at multiple frequencies.}
\label{fig:lightcurves}
\end{figure}

The early ($\Delta t\lesssim 40$ days) radio emission of SN\,2009bb
decays at both optically-thick and -thin frequencies
(Figure~\ref{fig:lightcurves}), in contrast to the predicted\cite{c98}
evolution for SNe Ibc in free expansion.  This indicates that another
process is responsible for powering the radio emission at early time:
a central engine.  In this context, the early evolution may be understood
if the geometry of the relativistic outflow was originally aspherical.
In this scenario, the relativistic ejecta are detached from the freely
expanding SN outflow and decelerate\cite{wax04} on a timescale,
$t_{\rm dec}\approx 1~(E/10^{49}~{\rm erg})(\dot{M}/10^{-5}~{\rm
M_{\odot}~yr^{-1})^{-1}}$ day, when $\Gamma\beta \lesssim 1$.  Here,
$\dot{M}=4\pi\rho v_w r^2$ is the mass loss rate of the progenitor
star, $\rho$ is the circumstellar density and $v_w$ is the velocity of
the progenitor wind. Based on the SSA model fits to the radio spectra,
we estimate\cite{cf06} the mass loss rate of the SN\,2009bb progenitor
to be $\dot{M}\approx 6.4 \times 10^{-7}
(\epsilon_e/\epsilon_B)^{-8/19} (\epsilon_B/0.33)^{-1} (f/0.5)^{-8/19}
(F_{\nu,p}/{\rm mJy})^{-4/19} (\nu_p/5~{\rm GHz})^2 (\Delta t/10~{\rm
days})^2~\rm M_{\odot}~yr^{-1}$ or $\dot{M}\approx 2.0\times
10^{-6}~\rm M_{\odot}~yr^{-1}$ for $f=0.5$ and
$\epsilon_e=\epsilon_B=0.33$ at $\Delta t\approx 20$ days.
 Here we have assumed a stellar wind density profile, $\rho\propto
 r^{-2}$, and a steady progenitor wind velocity of $v_w\approx
 10^3~\rm km~s^{-1}$ consistent with the observed\cite{cgv04}
 properties for local Wolf-Rayet stars.  For SN\,2009bb, we estimate
 that the outflow decelerates by $t_{\rm dec}\approx 6$ days and note
 that this estimate is independent of the partition fractions.
 Therefore the mean expansion velocity of $R/\Delta t\approx 0.9c$
 inferred at $\Delta t\approx 20$ days is a strict lower limit on the
 initial velocity of the outflow.  Thereafter
 the ejecta spread\cite{sph99} sideways, stalling the radial expansion
 as the outflow becomes spherical and causing the peak radio emission to
 decay.  The blastwave
 spherizes\cite{wax04} on a timescale, $t_{\rm sph}\approx
 6~(E/10^{49}~{\rm erg}) (\dot{M}/10^{-5}~{\rm
 M_{\odot}~yr^{-1}})^{-1}$ days.  By $\Delta t\approx 34$
 days, we estimate that the SN\,2009bb blastwave was spherical and the
 radio evolution thereafter is well described by a standard
 spherically symmetric model.

\section{X-ray Observations}

\subsection{Chandra data}

We observed SN\,2009bb with Chandra ACIS-S beginning at Apr 18.6 UT
($\Delta t\approx 31$ days) for 10 ksec under our approved Cycle 10
program to study the X-ray properties of radio-emitting Type Ibc
supernovae. We detect an X-ray source coincident with the optical and
radio counterparts at position $\alpha$(J2000)=\ra{10}{31}{33.86} and
$\delta$(J2000)=\dec{-39}{57}{29.4} ($\pm 0.4$ arcsec in each
coordinate). The data were reduced and the source counts were extracted
in the standard manner using the Chandra threads.  Within a 4.9 arcsec
aperture we extract 24 counts. After correcting for an inferred
foreground extinction of $E(B-V)\approx 0.5$ (see Ref.~\pcite{lsf+09})
corresponding\cite{ps95} to a column density, $n_H=2.92\times
10^{21}~\rm cm^{-2}$, we estimate an (absorbed) unabsorbed flux of
($F_X=1.5\pm 0.3\times 10^{14}~\rm erg~cm^{-2}~s^{-1}$) $F_X=2.3\pm
0.5\times 10^{-14}~\rm erg~cm^{-2}~s^{-1}$ (0.3-10 keV) assuming a
power-law spectrum with photon index, $\Gamma=2$.  At the distance of
NGC\,3278, the unabsorbed flux corresponds to an X-ray luminosity of
$L_X=4.4\pm 0.9\times 10^{39}~\rm erg~s^{-1}$.  The X-ray
emission from SN\,2009bb is therefore a factor of a few less luminous
than GRB\,980425 observed on a comparable timescale and at the high
end of the X-ray luminosities observed\cite{cf06} for other local SNe
Ibc.

\subsection{Swift X-ray Telescope Data}

We also observed SN\,2009bb with the {\it Swift} X-ray Telescope (XRT)
on several epochs spanning Mar 24.1 to Apr 18.7 UT.  Data were
retrieved from the HEASARC archive and reduction performed with the
{\tt xrtpipeline} script packaged within the HEAsoft software.  We
used the default grade selections and screening parameters to extract
measurements for the 0.3 -- 10 keV energy range.

We extracted the counts within a 47 arcsec (90\% PSF containment)
aperture centered on the optical SN position in each epoch.  Diffuse
emission from the host galaxy is clearly detected within the
extraction region.  We bootstrap the flux calibration to the Chandra
data, temporally coincident with our final XRT epoch.  Adopting the same 47
arcsec aperture for the Chandra data, we extract 233 counts after
subtracting off the background count rate.  Fitting the same absorbed
power-law model described above, we find an unabsorbed flux of
$F_X=3.0\pm 0.8\times 10^{-13}~\rm erg~cm^{-2}~s^{-1}$.  Therefore, within the
XRT PSF, the host galaxy emission dominates that of the SN by a factor
of $\sim 10$.  Since the fractional uncertainties of the XRT fluxes
preclude our ability to extract estimates of the SN flux, we adopt the
measurements as upper limits (see Table~\ref{tab:xray}).  Finally, we
note that a comparison of the first XRT upper limit and the measured
SN flux from the Chandra observation constrain the temporal evolution
of the SN X-ray flux to be no steeper than $F_X\propto t^{-1}$.

The flux ratio between radio and X-ray band can reveal the nature of
the X-ray emission which, in the case of SNe Ibc, may be dominated by
(i) thermal emission associated with the shock-heated ejecta, (ii)
non-thermal synchrotron emission from the shocked circumstellar
material, or (iii) inverse Compton up-scattering of optical photons by
radio-emitting electrons (see Ref.~\pcite{cf06} for a review).  In the
case of SN\,2009bb, at the epoch of the Chandra observation we measure
the spectral index between the radio ($\nu_R=8.46$ GHz) and X-ray band
($\nu_X=2\times 10^{17}$ Hz), to be $F_{\nu}\propto \nu^{-0.82\pm
0.02}$.  This is somewhat flatter than that expected by an
extrapolation of the radio synchrotron spectrum as $F_{\nu}\propto
\nu^{-1}$.  Given that the optical SN emission peaks on a similar
timescale, it is feasible that the X-ray emission is dominated by
inverse Compton emission. Alternatively, if the shock is cosmic ray
dominated the synchrotron spectrum flattens\cite{cf06} at higher
frequencies.  The nature of the X-ray emission will be discussed in a
subsequent paper.

\section{Searches for a Coincident High Energy Signal}

Massive star explosions can give rise to high energy emission at the
moment of explosion via two primary channels: (i) strong X-ray and/or
gamma-ray emission powered by ultra-relativistic ejecta as inferred
for GRBs, and (ii) weak X-ray emission produced by shock breakout in
ordinary non-relativistic supernova explosions (e.g., SN\,2008D).  In
order to search for a statistically significant association between an
optical SN and a high energy transient, an accurate estimate of the
explosion time is required.  SN\,2009bb is unique among local
optically-discovered SNe in that the optical data set is sufficient to
enable a reasonably tight constraint on the explosion date
(\S\ref{sec:t0}).  Along this line, we searched for high energy
counterparts between the dates Mar 18-20 UT and coincident with the SN
position.

Among the automatic triggered public bursts announcements made through
the Gamma-Ray Burst Coordination Network (GCN) from Swift, Fermi,
Integral, Agile, Suzaku and the Interplanetary Network (IPN), we note
just one marginally coincident burst. The weak Fermi Gamma-ray Burst
Monitor (GBM) burst GRB\,090320C was detected on Mar 20.045 UT at $\alpha\rm
(J2000)=07^{\rm h}13^{\rm m}$, $\delta\rm (J2000)=-43^{\rm o}18'$
(1$\sigma$ positional uncertainty of 15 deg; Ref.~\pcite{cha09}) and
lies $3.1\sigma$ away from the optical and radio SN\,2009bb positions.
With a duration of $T_{90}\approx 4$ sec, the flux of the burst is
$F_{\gamma}=3.9\times 10^{-7}~\rm erg~cm^{-2}~s^{-1}$ (8-1000 keV).
Given the large positional uncertainty due to the weakness of the
burst, we undertook a Monte Carlo experiment to calculate the
probability that GRB\,090320C was associated with SN\,2009bb.  We used
1000 iterations to test the frequency that a GRB with the properties
observed for GRB\,090320C could be recovered from a burst within
1$\sigma$ of the location of SN\,2009bb given the known detector
response.  From this experiment, we exclude the association of
GRB\,090320C and SN\,2009bb at the 99\% confidence level. Separately,
we note that the SN position was in the field-of-view of the GBM for
62 percent of our bracketed explosion date estimate.

We next searched for sub-threshold bursts in the Swift Burst Alert
Telescope (BAT) data, including those data from the BAT Slew Survey
(BATSS; Ref.~\pcite{gca+08}) for the same range of dates. The SN
position was within the 50 deg field-of-view of the BAT for just 7
percent of this time.  With an accuracy of $\sim 2$ arcmin, no
sub-threshold ($F_{\gamma}\gtrsim 5.5\sigma$ above background)
triggers were detected at a position coincident with SN\,2009bb.
Moreover, in the BATSS data set we find no statistically significant
($F_{\gamma}\gtrsim 3.5\sigma$) bursts during the spacecraft slews
over this timescale.  We also note that GBM burst 090320C was not in
the field-of-view of the BAT at the time of the Fermi trigger.

By combining the monitoring times of BAT and GBM and accounting for
the 9772 seconds when the SN position was monitored by both satellites
simultaneously, we estimate a combined monitoring time of 65 percent.
Furthermore, both GBM and BAT are included within the all-sky
Interplanetary Network (IPN; ) of gamma-ray satellites which offers
100\% temporal coverage to SN\,2009bb over the bracketed explosion
date estimate.  Based on the detection efficiency of IPN as reported
in Ref.~\pcite{hcm+09}, we place an upper limit on the detectable
gamma-ray fluence from SN\,2009bb of $F_{\gamma}\lesssim 5\times
10^{-6}~\rm erg~cm^{-2}$ (band, 25-150 keV).  At a distance of
$d\approx 40$ Mpc, this corresponds to an isotropic-equivalent energy
release of $E_{\gamma}\lesssim 10^{48}$ erg which is a factor of two
higher than that observed\cite{paa+00} from GRB\,980425.

The lack of a detected gamma-ray counterpart can be attributed to
several factors.  For example, if the GRB jet is not directed towards
our line-of-sight, relativistic beaming inhibits the detection of
gamma-rays.  Most importantly, as seen in the case of the nearest
GRBs, the gamma-rays can be exceedingly weak and the isotropic
gamma-ray energy yield dwarfed by that of the blastwave kinetic energy
(e.g., GRB\,980425; Ref.~\pcite{paa+00}).  Along this line we note
that the rate of such intrinsically weak GRBs at $z\lesssim 0.1$ is
3-10 times higher\cite{skn+06} than that of classic GRBs discovered
preferentially at larger redshifts.

\section{Predictions for the Rate of Engine-driven SNe from
Long-Wavelength Surveys}

Local ($d\lesssim 200$ Mpc) SNe Ibc are currently
discovered\footnote{http://www.cfa.harvard.edu/iau/lists/Supernovae.html}
at a rate of $\sim 25~\rm yr^{-1}$ by the ongoing efforts of amateur
astronomers and automated searches (e.g., Lick Observatory Supernova
Search, CHASE) that monitor weekly a sample of nearby galaxies
searching for new or transient objects.  With the advent of wide-field
optical surveys, specifically The Palomar Transient Factory (PTF;
Ref.~\pcite{lkd+09}) and Pan-STARRS\cite{kab+02}, this sample will be
supplemented by similarly nearby and young SNe Ibc at a rate of $\sim
50~\rm yr^{-1}$ (PTF) and $\sim 25~\rm yr^{-1}$ (Pan-STARRS) over the
next 3-5 yrs given the volumetric rate\cite{cet99,dsr+04} of SNe Ibc
of $1.7\times 10^{4}~\rm Gpc^{-3} yr^{-1}$.

Through prompt radio observations of the 100 SNe $\rm yr^{-1}$
discovered by these combined efforts, engine-driven events like
SN\,2009bb can be clearly identified by their luminous and
low-frequency radio spectra.  The current sensitivity of the Very
Large Array ($3\sigma\approx 0.1$ mJy in 15 min) implies that the
radio luminosity light-curve of SN\,2009bb extending to $\sim 100$
days could be detected with S/N $\gtrsim 5$ out to a distance of 130
Mpc.  The increase in sensitivity of the Expanded-VLA (expected 2010;
Ref.~\pcite{pnb04}) will enable such events to be detected
significantly further, to 400 Mpc, well matched to the volumes probed
by the optical surveys, and requiring just 25 hrs of observing time
per year.  Given our estimated fraction of engine-driven events,
$0.7^{+1.6}_{-0.6}\%$, we anticipate that $\sim 1$ new relativistic SN
will be both discovered and identified each year.  This is a factor of
three higher than the current discovery rate\cite{skn+06} of nearby
GRBs with gamma-ray satellites, demonstrating that long-wavelength
surveys will ultimately become powerful tools in the discovery of such
events.  In fact, over the next decade, the ``deep-wide-fast" optical
survey planned for the Large Synoptic Survey Telescope (LSST;
Ref.~\pcite{ita+08}) will reveal $\sim 500$ SNe Ibc each year within
400 Mpc, of which up to $\sim 10$ will be engine-driven.

\bibliographystyle{nature-pap}

\clearpage

\begin{table}
\begin{center}
\begin{tabular}{>{\small}c >{\small}c >{\small}c >{\small}c >{\small}c >{\small}c}
\hline\hline
Date & $F_{\nu,1.43~\rm GHz}$ & $F_{\nu,4.86~\rm GHz}$ &  $F_{\nu,8.46~\rm GHz}$ & $F_{\nu,22.5~\rm GHz}$  &  VLA \\
(UT) & (mJy) & (mJy) & (mJy) & (mJy) & Config. \\\hline
Apr 5.2  & \nodata & \nodata & $24.681\pm 0.066$ & \nodata & B \\
Apr 8.2  & $1.308\pm 0.099$ & $12.863\pm 0.161$ & $17.568\pm 0.088$ & $5.204\pm 0.191$ & B \\
Apr 13.2 & \nodata & \nodata & $16.349\pm 0.107$ & \nodata & B \\
Apr 15.1 & \nodata & $9.629\pm 0.181$ & $13.812\pm 0.114$ & $5.170\pm 0.394$ & B \\
Apr 23.2 & $1.141\pm 0.159$ &  $7.601\pm 0.226$ &  $8.881\pm 0.121$ & \nodata & B \\
Apr 29.1 & $2.963\pm 0.348$ & $7.632\pm 0.202$ & $7.714\pm 0.095$ & $1.796\pm 0.240$ & B \\
May 3.1  & $3.406\pm 0.122$ & $9.680\pm 0.165$ & $8.482\pm 0.098$ & \nodata & B \\
May 10.1 & $4.939\pm 0.132$ & $12.215\pm 0.196$ & $6.824\pm 0.102$ & \nodata & B \\
May 13.0 & $5.836\pm 0.136$ & $9.727\pm 0.508$ &  $6.327\pm 0.151$ & \nodata & B \\
May 20.1 & $6.903\pm 0.180$ & $8.542\pm 0.172$ &  $3.294\pm 0.118$ &  \nodata & B \\
May 27.0 & $6.702\pm 0.351$  & $8.568\pm 0.080$ &  $4.204\pm 0.060$ & \nodata & CnB \\
Jun 6.0 & $9.465\pm 0.186$ & $5.877\pm 0.110$ & $3.203\pm 0.074$ & \nodata & CnB \\
Jun 17.0 & $12.133\pm 0.249$ & $4.220\pm 0.120$ & $2.392\pm 0.082$ & \nodata & CnB \\
Jun 26.0 & $12.189\pm 0.522$ & $3.556\pm 0.350$ & $1.903\pm 0.548$ & \nodata & C \\
Jul 18.9 & $11.230\pm 0.356$ & $2.510\pm 0.161$ & $1.032\pm 0.104$ & \nodata & C \\  
Aug 11.8 & $8.687\pm 0.341$ & $2.477\pm 0.155$ & $1.084\pm 0.091$ & \nodata & C \\
\hline
\end{tabular}
\caption[Figure~1]{\small Very Large Array observations of SN\,2009bb.}   
\label{tab:vla} 
\end{center}
\end{table}

\begin{table}
\begin{center}
\begin{tabular}{>{\small}c >{\small}c >{\small}c >{\small}c >{\small}c >{\small}c}
\hline\hline
Date & Frequency & Flux Density & Synthesized Beam Size \\
(UT) & (MHz) & (mJy) & (arcsec)\\\hline
May 31.5 & 1287 & $4.6\pm 0.3$ & 5x2 \\
Jun 10.6 & 617 & $3.0\pm 0.3$ & 10x4 \\
Aug 8.3 &  617 & $7.9\pm 0.3$ & 10x4 \\
Aug 10.4 & 332 & $3.3\pm 0.7$ & 13x7\\
\hline
\end{tabular}
\caption[]{\small GMRT observations of SN\,2009bb.}
\label{tab:gmrt} 
\end{center}
\end{table}

\begin{table}
\begin{center}
\begin{tabular}{>{\small}c >{\small}c >{\small}c >{\small}c >{\small}c >{\small}
c}
\hline\hline
Date & Exposure & Flux & Mission \\
(UT) & (sec) & ($\rm erg~cm^{-2}~s^{-1}$) &  \\\hline
Mar 24.1 & 3964 & $\lesssim 1.3\times 10^{-13}$ & {\it Swift}/XRT \\ 
Apr 6.2 & 9829 & $\lesssim 1.7\times 10^{-13}$ & {\it Swift}/XRT \\ 
Apr 10.5 & 6720 & $\lesssim 2.5\times 10^{-13}$ & {\it Swift}/XRT \\
Apr 18.7 & 12752 & $\lesssim 3.0\times 10^{-13}$ & {\it Swift}/XRT \\
Apr 18.6 & 9914 & $2.3\pm 0.5\times 10^{-14}$ & Chandra \\
\hline
\end{tabular}
\caption[Figure~1]{\small X-ray observations of SN\,2009bb. All measurements are 0.3--10 keV and unabsorbed.  Upper limits are dominated by diffuse host galaxy emission.}   
\label{tab:xray} 
\end{center}
\end{table}

\clearpage

\clearpage

\end{document}